%% file: manuscript.tex
\documentstyle[twocolumn,aps,prb]{revtex}

\input psfig2

\def\al2o3{Al$_2$O$_3$}
\def\aal2o3{$\alpha$-Al$_2$O$_3$}

\begin{document}
\draft


\title{The $\Sigma 13$ $(10\bar{1}4)$ twin in \aal2o3: A 
model for a  general grain boundary}
\author{Stefano Fabris and Christian Els\"asser} 
\address{Max-Planck-Institut f\"ur Metallforschung, Seestrasse 92,
D-70174 Stuttgart, Germany}
\date{\today} 
\maketitle
\begin{center}
\begin{abstract}
 
 The atomistic structure and energetics of the $\Sigma 13$ $(10\bar{1}4)$
 symmetrical tilt grain boundary in $\alpha$-Al$_2$O$_3$ are studied by
 first-principles calculations based on the local-density-functional
 theory with a mixed-basis pseudopotential method.  Three configurations,
 stable with respect to intergranular cleavage, are identified: one
 Al-terminated glide-mirror twin boundary, and two O-terminated twin
 boundaries, with glide-mirror and two-fold screw-rotation symmetries,
 respectively. Their relative energetics as a function of axial grain
 separation are described, and the local electronic structure and bonding
 are analysed. The Al-terminated variant is predicted to be the most
 stable one, confirming previous empirical calculations, but in contrast
 with high-resolution transmission electron microscopy observations on
 high-purity diffusion-bonded bicrystals, which resulted in an
 O-terminated structure.  An explanation of this discrepancy is proposed,
 based on the different relative energetics of the internal interfaces
 with respect to the free surfaces.

\end{abstract}
\pacs{61.72.Mm,61.72.Bb,73.20.At,71.15.Ap}
\end{center}

\narrowtext



\section{Introduction}

 Alumina (\aal2o3)
 is widely used in polycrystalline forms. Small amount
 of various dopant elements (typically Mg, Ca, Si, and Ti) are added in the
 synthesis of alumina polycrystals, either to promote sintering,
 or to obtain empirically optimized microstructures (grain size
 and shape distributions) for technological applications.~\cite{KingeryAdv}
 Due to the very small solubility limit in bulk for many impurities
 (typical values are 10 ppm for yttrium,~\cite{Cawley86} 30 ppm for
 calcium,~\cite{Bae93jamcer} 300 ppm for silicon~\cite{Bae93jamcer}), they
 segregate to grain boundaries (GB).~\cite{Li84adv} The segregation of
 different species at grain boundaries drastically modifies the properties
 of alumina, in either beneficial or undesired manner. For instance, the
 isovalent cations Y and La reduce the mechanical creep rate of
 polycrystalline alumina by 2$-$3 orders of
 magnitude,~\cite{Lartigue93,Cho97} Si and Ca promote exaggerated grain
 growth,~\cite{Bae93jamcer,Bae93jmat} and Ca may affect the fracture
 properties, decreasing the strength and toughness of the
 material.~\cite{Funkenbusch75,Jupp80} From a fundamental aspect of 
 science, a key issue in alumina systems is therefore the microscopic
 understanding of internal interfaces and of grain-boundary impurity
 segregation.

%
%


 This study is part of an extended project that aims to correlate the
 macroscopic structural properties of alumina to its microscopic
 structure. The project combines empirical atomistic modelling and
 first-principles electronic-structure theory with experimental
 high-resolution transmission electron microscopy (HRTEM) and with 
 analytical electron microscopy (AEM), in particular spatially resolved 
 electron energy-loss
 near-edge spectroscopy (ELNES). In this framework a set of
 four grain boundaries were selected for a fundamental theoretical study of
 atomistic and electronic structures: the basal $\Sigma 3$ (0001), the
 rhombohedral $\Sigma 7$ ($\bar{1}012$), the $\Sigma 13$ ($10\bar{1}4$),
 and the prismatic $\Sigma 3$ ($10\bar{1}0$) twin boundaries. This set
 provides a couple of mutually orthogonal twin interfaces, 
 basal and prismatic,
 and a couple of nearly orthogonal ones, the $\Sigma 7$ and the $\Sigma
 13$.

 The basal and rhombohedral twin interfaces have recently been studied in
 detail by means of ab-initio local-density-functional theory (LDFT) and
 empirical shell-model calculations.~\cite{Marinop00,Marinop01,Elsaesser01}
 The theoretical analysis of the interface structure for the rhombohedral
 twin has been confirmed quantitatively by both HRTEM and ELNES
 experiments,~\cite{Nufer01} validating the accuracy and predictive power
 of the theoretical approach, which is employed as well in the present
 work dealing with the $\Sigma 13$ twin interface. Results for the 
 prismatic twin interface will be reported separately~\cite{prismtwin}.

 Several studies indicated that the segregation of impurities like Ca or Y
 to GB is more favourable for general, disordered interfaces than for
 special ones with high structural order and energetic
 stability.~\cite{Bouchet93,Swiatnicki95} This is commonly rationalized by
 the large mismatch in ionic radii between the segregated species
 (r$_i$=0.94 \AA\ for Ca$^{2+}$ and r$_i$=0.88 \AA\ for Y$^{3+}$,
 Ref.~\onlinecite{Kittel}) and the host-lattice element Al$^{3+}$
 (r$_i$=0.45 \AA, Ref.~\onlinecite{Kittel}).  General boundaries provide
 more excess volume at the interface, and the oversized Ca and Y cations
 fit more easily here than at special GB, where the interfacial atomic
 environment is highly ordered and compact. As an example, the impurity
 level in the nearly bulk-like rhombohedral and prismatic twin boundaries
 is below the detection limit of 0.3 atoms/nm$^2$.~\cite{NuferTh} There are
 indications that the basal plane may have an atomic environment which is
 slightly more favourable to segregation.~\cite{NuferTh,Kaplan95,Brydson98}
 Compelling experimental evidence for segregation to a twin boundary was
 observed for the case of the $\Sigma 13$ ($10\bar{1}4$) twin
 boundary.~\cite{Hoche96}

 Among the four selected GB, the interfacial structure of the $\Sigma 13$
 boundary is the most different from the bulk atomic environment of
 \aal2o3. Hence it is anticipated to have a rather high interface
 energy, and it seems to be best suited and most promising to address the
 issue of segregation of cation impurities to GB.  Its interfacial
 structure 
 appears 
 sufficiently disordered and open to accept likely 
 segregated impurities, thus
 being representative for the atomic environment in a general GB. At the
 same time it is sufficiently ordered, with short lateral periodicity along
 the interface, that it can be described with some tens of atoms.  Hence it
 is accessible via first-principles LDFT calculations. In the present work
 we consider the $\Sigma 13$ twin boundary as a representative case to
 study the effect of segregation on the mechanical and electronic
 properties of alumina interfaces. We describe the properties of the pure
 $\Sigma 13$ twin interface in this paper, and the modifications induced by
 the segregated atoms in a forthcoming one.


 Theoretical investigations preceding ours of the same $\Sigma 13$ twin
 have been undertaken by Finnis and coworkers.  To our best knowledge,
 results have been published only for empirical shell-model
 calculations,~\cite{Hoche94a,Kenway94,Exner96} but not for ab-initio LDFT
 calculations.~\cite{Huang97unp}

 This paper is arranged as follows: A short review of previous studies on
 the $\Sigma 13$ twin boundary is presented in Section~\ref{prevstud}. A
 brief desciption of the employed computational methods is given in
 Sec.~\ref{compdet}. The results of our investigation, namely the
 mechanical stability, the relative energetics, and the electronic
 structure of different metastable structural variants for the $\Sigma 13$
 interface are reported and discussed in Sec.~\ref{results}. Final remarks
 and a summary are made in Section~\ref{concl}.

\section{Previous studies on the $\Sigma 13$ twin boundary}
\label{prevstud}
 
\begin{figure}
\caption{Corundum structure of \aal2o3 viewed from the [$ 1\bar{2}10$]
direction. Oxygen and aluminium atoms are represented by light-gray and
black circles, respectively. Dashed lines mark the ($10\bar{1}4$) planes
leading to stoichiometric $\Sigma 13$ interfaces with Al (B) and O (C)
terminations, respectively.}
\label{cryst}
\centerline{\psfig{file=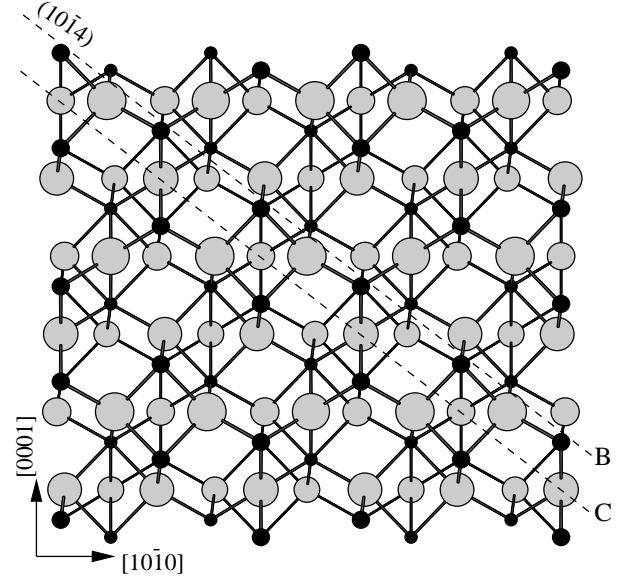,width=8cm,angle=0}} 
\end{figure}

  Figure~\ref{cryst} shows the crystal structure of \aal2o3 (corundum,
 space group $R\bar{3}c$), projected on the ($\bar{1}2\bar{1}0$) plane:
 light and dark circles represent oxygen and aluminium ions, respectively.
 The structure can be described as a nearly ideally close packed
 hexagonal oxygen sublattice with 2/3 of
 the available octahedral interstices filled by Al cations. The remaining
 octahedral sites are empty, and the partial filling results in small
 internal atomic displacements in both sublattices, which are visible in
 Figure~\ref{cryst}.  The $\Sigma 13$ ($10\bar{1}4$) twin interface is
 formed by two grains with a relative misorientation given by the planes
 ($10\bar{1}4$)$||$($10\bar{1}4$) and the directions
 [$1\bar{2}10$]$||$[$\bar{1}2\bar{1}0$].

 High-resolution transmission electron micrographs of high-purity
 diffusion-bonded bi\-crys\-tals containing this interface~\cite{Hoche94a}
 revealed two distinct boundary structures. The images differed by the
 sequence of intensity maxima across the boundary: one showed a continuous
 (C) and the other a broken (B) pattern of spots across the boundary.  The
 C interface was found in the bicrystal bonded without further treatment,
 besides sufficient cleaning of the free surfaces, while the B and C
 interfaces coexisted when a metallic Al layer was deposited on the free
 surfaces after the cleaning and before the diffusion-bonding process.

 Atomistic modelling based on empirical 
 classical potentials~\cite{Hoche94a,Kenway94}
 also identified two boundary structures, in which the two grains were
 either Al or O terminated. They predicted the former to be more stable
 than the latter. The calculated interface energies for the two relaxed
 boundaries were 1.7 J m$^{-2}$ (Al-terminated) and 2.1 J m$^{-2}$
 (O-terminated), respectively.

 The details of these interface structures were not directly accessible
 from the HRTEM images alone. The combination of atomistic modelling and
 image simulations allowed an assignment of the B and C patterns to cation
 and anion terminated interfaces, respectively.~\cite{Hoche94a} The
 corresponding terminating planes in the perfect corundum structure of
 Figure~\ref{cryst} are marked by the dashed lines labelled B and C.


 H\"oche and R\"uhle studied the segregation of calcium to the $\Sigma 13$
 boundary and the resulting structural modifications induced on the B and C
 structures.\cite{Hoche96} A layer of metallic calcium, 3 nm thick, was
 deposited on one of the ($10\bar{1}4$) free surfaces after cleaning and
 before diffusion-bonding the bicrystal. After the bonding procedure,
 2.4$\pm$0.9 Ca ions per nm$^2$ were detected at the
 interface,~\cite{HocheTh} therefore proving that Ca segregates to this
 grain boundary.  Again, as in the Al-doped case, two distinct structures
 coexisted: one variant showed little differences to the C structure. The
 other was qualitatively similar to the B structure but with significant
 quantitative differences.  In this case the comparison of the HRTEM with
 theoretical calculations was not possible because atomistic calculations
 of cation segregation at alumina grain boundaries were not available.

 Such results raise fundamental questions on the formation mechanism and
 kinetics of interfaces. In the specific case of the $\Sigma 13$ twin
 boundary, the most common structure observed in diffusion-bonded
 bicrystals (O-terminated) is apparently 
 not the one with the minimum energy predicted
 by the theoretical calculations (Al-terminated).  
 Hence, is the $\Sigma 13$ twin
 boundary a case in which
 empirical atomistic model potentials reach their limit of predictive
 power? 
 This concern is assessed by employing the 
 non-empirical LDFT approach for the theoretical investigation.
 Or is the $\Sigma 13$ twin
 boundary rather a case where diffusion-bonding of the lowest-energy
 surfaces does not immediately result in the lowest-energy interface? 
 This possibility is investigated by mapping theoretically the relative 
 energetics along configuration paths for both bonding the most stable 
 surfaces together and pulling the most stable interface apart.
 The paper provides a precise description of the interfacial atomic
 environment and electronic structure of the $\Sigma 13$ twin boundary, and
 it proposes an explanation of the difference between experimental HRTEM
 observations and theoretical predictions.




\section{Computational methods}
\label{compdet}

 In this work we used the same computational approach as previously
 for the rhombohedral twin boundary,~\cite{Marinop00} the basal
 twin boundary,~\cite{Marinop01} and basal-plane stacking
 faults,~\cite{Marinop01stack} namely atomistic modelling with empirical
 interatomic potentials and ab-initio LDFT calculations.

 The empirical potentials which were used in the atomistic modelling of
 \aal2o3, were taken from Lewis and Catlow~\cite{Lewis85}. 
 They incorporate the
 long-range Coulomb interaction between all Al cations and O anions, the
 short-range ion-overlap repulsion, the weak Van-der-Waals attraction
 between the anions, and a shell-model dipole polarizability of the
 anions. The cations were treated as non polarizable and mutually
 interacting via Coulomb repulsion only.  The advantage of the empirical
 potentials with respect to LDFT is their superior computational
 simplicity, the disadvantage is their limited reliability for predictions
 of material-specific properties.  The comparison of the two
 approaches~\cite{Marinop01,Marinop01stack} for basal-plane
 twin-boundary structures and stacking faults in \aal2o3 showed that the
 empirical potentials can provide quantitatively not very accurate
 interface energies, but qualitatively correct energy hyperfaces in
 structural parameter spaces. Atomistic arrangements in metastable
 structures obtained by both approaches were found to be very similar,
 making the empirical potentials very useful for identifying relevant
 initial grain-boundary configurations, that can then be structurally
 optimized with the LDFT.

%
%

 The LDFT calculations were made by means of the mixed-basis
 pseudopotential technique for the computation of electronic structures,
 total energies, and forces in crystalline materials. The interfaces were
 described by supercells with periodic boundary conditions.  The
 computational details, namely the norm-conserving ionic pseudopotentials
 for Al and O, the mixed basis for the representation of the valence-band
 and conduction-band Bloch states in \aal2o3, the discrete Brillouin-zone
 integration meshes etc. were the same as, or equivalent to the ones used
 in our previous investigations.~\cite{Marinop00,Marinop01,Elsaesser01}
 For the optimization of the interface structures, excess expansions of the
 interfaces were obtained via total-energy minimization with respect to the
 axial supercell length. Simultaneously, to account for lateral grain
 translations as well, all atomic positions in the supercells were
 statically relaxed according to the acting forces.

 Numerous published studies of various materials' structures and properties
 from the Stuttgart groups and others 
 involved in developments of the mixed-basis
 pseudopotential
 approach~\cite{Elsaesser90,Ho92,ElsaesserTh,MeyerTh,MeyerUnp} have
 documented the accuracy, efficiency and predictive power of this
 density-functional technique. Specifically for \aal2o3, bulk structural
 parameters of corundum were compared to results from other LDFT methods
 and from experiment in Table 1 of Ref.~\onlinecite{Marinop00}, and the
 predictive power of the method for grain-boundary structures was
 demonstrated in Ref.~\onlinecite{Marinop01stack}. 

 The unit cells used to model the $\Sigma 13$ twin interface were
 constructed by cutting the corundum structure of \aal2o3 with a
 ($10\bar{1}4$) plane and rotating one half of the crystal with respect to
 the other by 180$^\circ$, around the interface-plane normal vector {\bf
 n}$_{\rm GB}$=[$50\bar{5}4$] (symmetrical twist). Note that in hexagonal
 coordinates the [$50\bar{5}4$] direction is orthogonal to the
 ($10\bar{1}4$) plane. The rotation axis {\bf n}$_{\rm GB}$ contains a
 center of inversion symmetry for the original corundum structure,
 therefore the resulting bicrystal is mirror-symmetric.

 This procedure was applied to the cutting planes B and C marked by the
 dashed lines in Figure~\ref{cryst}, leading to the starting configurations
 for two stoichiometric variants of the grain boundary. The computational
 cell was then obtained by cutting a slab centered on the grain boundary,
 and repeating it with periodic boundary conditions. The slab-edges
 identifying the shape of the orthorhombic supercell are {\bf e}$_1$ $\|$
 [$20\bar{2}\bar{1}$], {\bf e}$_2$ $\|$ [$1\bar{2}10$], and {\bf e}$_3$
 $\|$ [$50\bar{5}4$].

 The size of the slab along {\bf e}$_3$ was selected so that it contained
 two equivalent interfaces, one in the middle, already described, and one
 at the edges, arising as a consequence of the periodic boundary
 conditions. A careful choice of the supercell size along {\bf e}$_3$ was
 necessary in order to minimize the spurious interaction between the two
 equivalent boundaries. The convergence of the results against the
 interface-interface distance was checked by considering 3 supercells, with
 10, 15, and 20 atomic layers between the boundaries, corresponding to 40-,
 60-, and 80-atom supercells. The results for the small cell were already
 converged in the case of the cation terminated interface while the 60-atom
 cell was necessary for the anion terminated one. We will further discuss
 this issue in the following.

\section{Results and discussion}
\label{results}

\subsection{Lateral translation state}

 The twin-boundary equilibrium configurations were found with increasing
 degrees of accuracy, by combining empirical atomistic and first-principles
 calculations. Previous studies showed that the empirical shell-model
 potential provides the correct qualitative description of the energetics
 and structure of twin boundaries.~\cite{Marinop00,Marinop01} This makes
 the shell-model suitable for a first exploration of possible metastable
 configurations.

 Using this model and the computer program GULP,~\cite{GaleGULP} we
 calculated the bicrystal energetics as a function of the relative
 displacement of one grain with respect to the other along the interface
 plane. The relative displacement is defined by the lateral translation
 vector $T_1{\bf e}_1+T_2{\bf e}_2$, where $T_1$ and $T_2$ are fractional
 coordinates with respect to the lattice vectors ${\bf e}_1$ and ${\bf
 e}_2$. The interfacial symmetry confines the non-equivalent part of the
 energy surface to a quarter of the lateral repeat unit cell, limited by
 the interval between 0 and 0.5 for both $T_1$ and $T_2$.

 The starting boundary structures, corresponding to ($T_1,T_2$)=(0,0), were
 the mirror-symmetric interfaces obtained as described in the previous
 section. It is clear that these structures are highly unstable because
 equal charges on opposite sides of the boundary face each other at very
 close distances. This is the reason why atomic relaxation was not
 considered in this preliminary analysis of the translation state. The
 shell-model potentials are not suitable to describe such highly
 non-equilibrium structures and would predict large unrealistic
 relaxations. Only the metastable structures identified with the static
 shell-model calculations were relaxed subsequently, according to both
 empirical shell-model and first-principles forces on the individual atoms.

\subsubsection{Cation terminated interface}

\begin{figure}
\caption{Shell-model total energy (Ry/supercell) for the cation
terminated interfaces as a function of the lateral translation state
$T_1{\bf e}_1+T_2{\bf e}_2$, with {\bf e}$_1$=[$20\bar{2}\bar{1}$] and
{\bf e}$_2$=[$1\bar{2}10$].}
\label{cshell}
 \centerline{\psfig{file=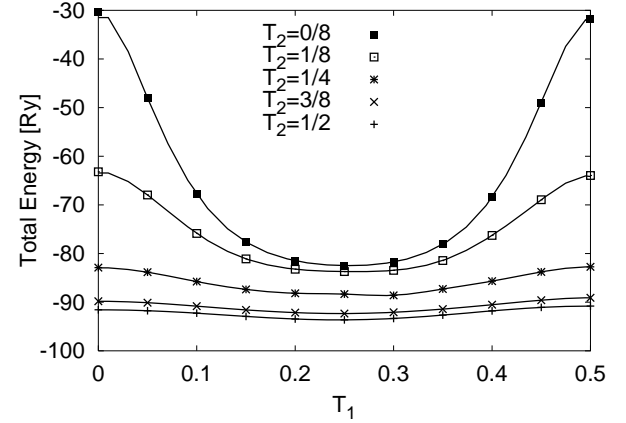,width=8cm,angle=-90}} 
\end{figure}

 The cation terminated interface was obtained by cutting the corundum
 structure along line B in Figure~\ref{cryst} and building the supercell as
 described above. The energy surface, governing the inter-granular lateral
 translation state, as a function of ($T_1,T_2$) is shown in
 Figure~\ref{cshell}.

\begin{figure}
\caption{LDFT result for the relaxed atomic structure of the G(Al)
interface viewed along the [$1\bar{2}10$] direction. 
The atoms are represented as in
Figure~\ref{cryst}, and the labels on the right-hand side identify the
atomic layers.  The dashed box indicates the axial size of the 60-atom supercell
in the [$50\bar{5}4$] direction, and both the size and lateral
translation-state in the [$20\bar{2}\bar{1}$] direction. 
The supercell
contains two equivalent horizontal interfaces, one in the middle (marked
by the dotted line), and one
both at the top and bottom ends (marked by the dashed line segments)
arising from the periodic boundary conditions
in the [$50\bar{5}4$] direction. 
}
\label{cfig}
 \centerline{\psfig{file=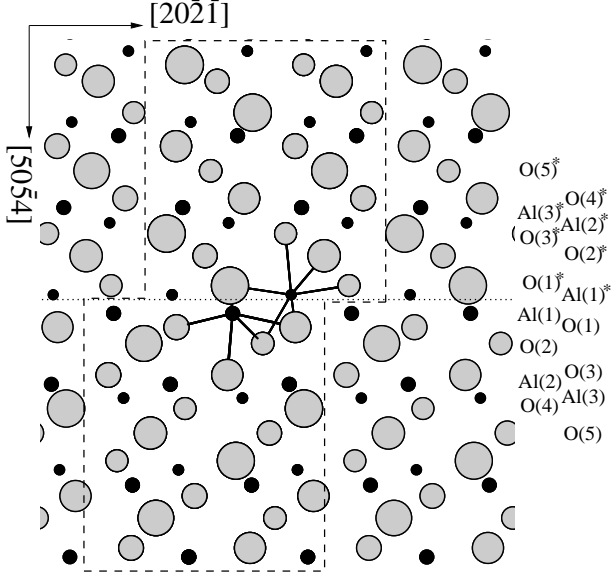,width=8.0cm,angle=0}} 
\end{figure}

 The starting configuration ($T_1,T_2$)=(0,0) is expected to be unstable
 because it faces equal charges very close to each other, on opposite sides
 of the boundary. The shell-model predicts the highest energy for this
 configuration. The total energy is minimized by the lateral translation
 state ($T_1,T_2$)=(0.25,0.5), resulting in an Al terminated structure with
 glide-mirror symmetry with respect to ($10\bar{1}4$).  This configuration,
 which we will indicate as G(Al), maximizes the inter-granular distance
 between equal charges and minimizes that one between opposite
 charges. Alternating triplets of oxygen atoms at the interface
 (O(1)-O(2)-O(3) and O(1)$^{\ast}$-O(2)$^{\ast}$-O(3)$^{\ast}$ in
 Figure~\ref{cfig}) characterize this boundary.

 The structure was then relaxed according to the forces from the 
 shell model as well as from 
 the LDFT. The results were qualitatively equivalent: the stable translation
 state was the same, and internal relaxations involved only ions in the
 neighbourhood of the interface, namely the atomic layers from O(2) to
 O(2).$^{\ast}$ The LDFT relaxed structure is shown in
 Figure~\ref{cfig}. It has the same features as the cation-terminated
 interface identified in Ref.~\onlinecite{Hoche94a}: an HRTEM setting for
 this interface showing the holes as intensity maxima would produce a
 broken pattern of white spots across the boundary.

 In order to check the adequacy of the supercell size perpendicular to the
 interface, the relaxation procedure was made for 40- and 60-atom
 supercells. The small difference between the interface energies ($< 4 \%$)
 of the two supercells showed that convergence was already achieved with
 the small cell.


\subsubsection{Anion terminated interfaces}

\begin{figure}
\caption{Shell-model total energy for the anion terminated interfaces as
a function of the lateral translation state $T_1{\bf e}_1+T_2{\bf e}_2$,
where {\bf e}$_1$=[$20\bar{2}\bar{1}$] and {\bf e}$_2$=[$1\bar{2}10$].}
\label{ashell}
 \centerline{\psfig{file=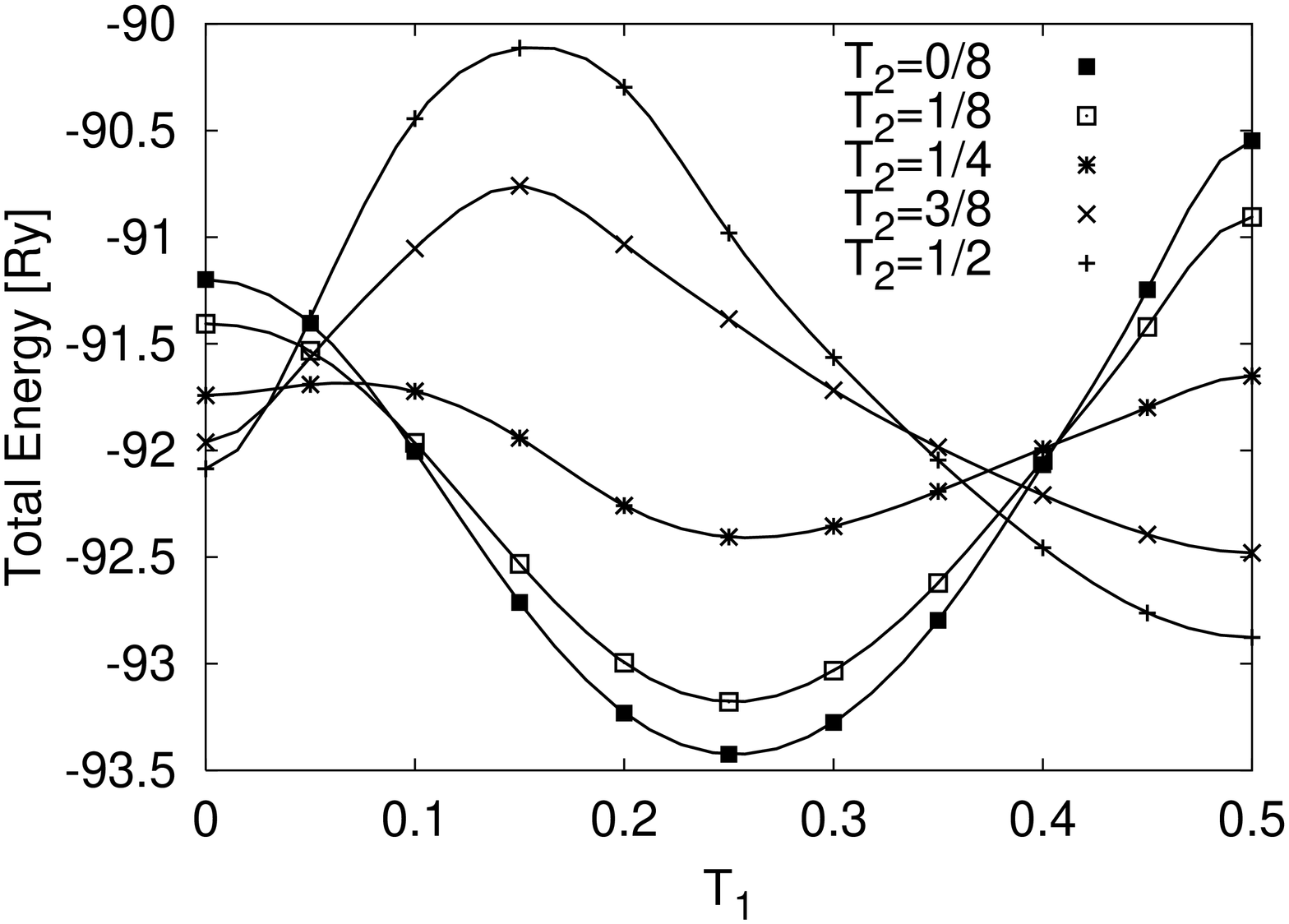,width=8cm,angle=-90}} 
\end{figure}

 A similar procedure was carried out for the anion terminated interface,
 which was obtained by cutting the single crystal along line C in
 Figure~\ref{cryst}. Note that in this case the interface plane coincides
 with an atomic oxygen layer. The total energy as a function of the lateral
 translation state ($T_1,T_2$) is shown in Figure~\ref{ashell} and predicts
 three metastable configurations.


 The structure with the lowest energy has a translation state of
 ($T_1,T_2$)=(0.25,0), and corresponds to the Al-terminated boundary
 already described above. This is an important feature of this boundary:
 the same Al-terminated interface may be obtained either by cutting the
 single crystal along B (Figure~\ref{cryst}), rotating one grain around
 {\bf n}$_{\rm GB}$ by 180$^\circ$ and imposing the lateral translation
 state ($T_1,T_2$)=(0.25,0.5), or by cutting along C, rotating one grain
 around {\bf n}$_{\rm GB}$ by 180$^\circ$ and imposing the lateral
 translation state ($T_1,T_2$)=(0.25,0). In the latter case the initial
 boundary was on an O layer, but the glide-mirror symmetry operation
 reconstructs a bulk-like environment near the atomic layer at C, and
 shifts the boundary by two atomic layers from O termination (C) to Al
 termination (B).

\begin{figure}
\caption{LDFT result for the relaxed atomic structure of the G(O)
interface viewed along the [$1\bar{2}10$] direction. Atoms, labels, and
symbols as in Figure~\ref{cfig}.}
\label{afig}
 \centerline{\psfig{file=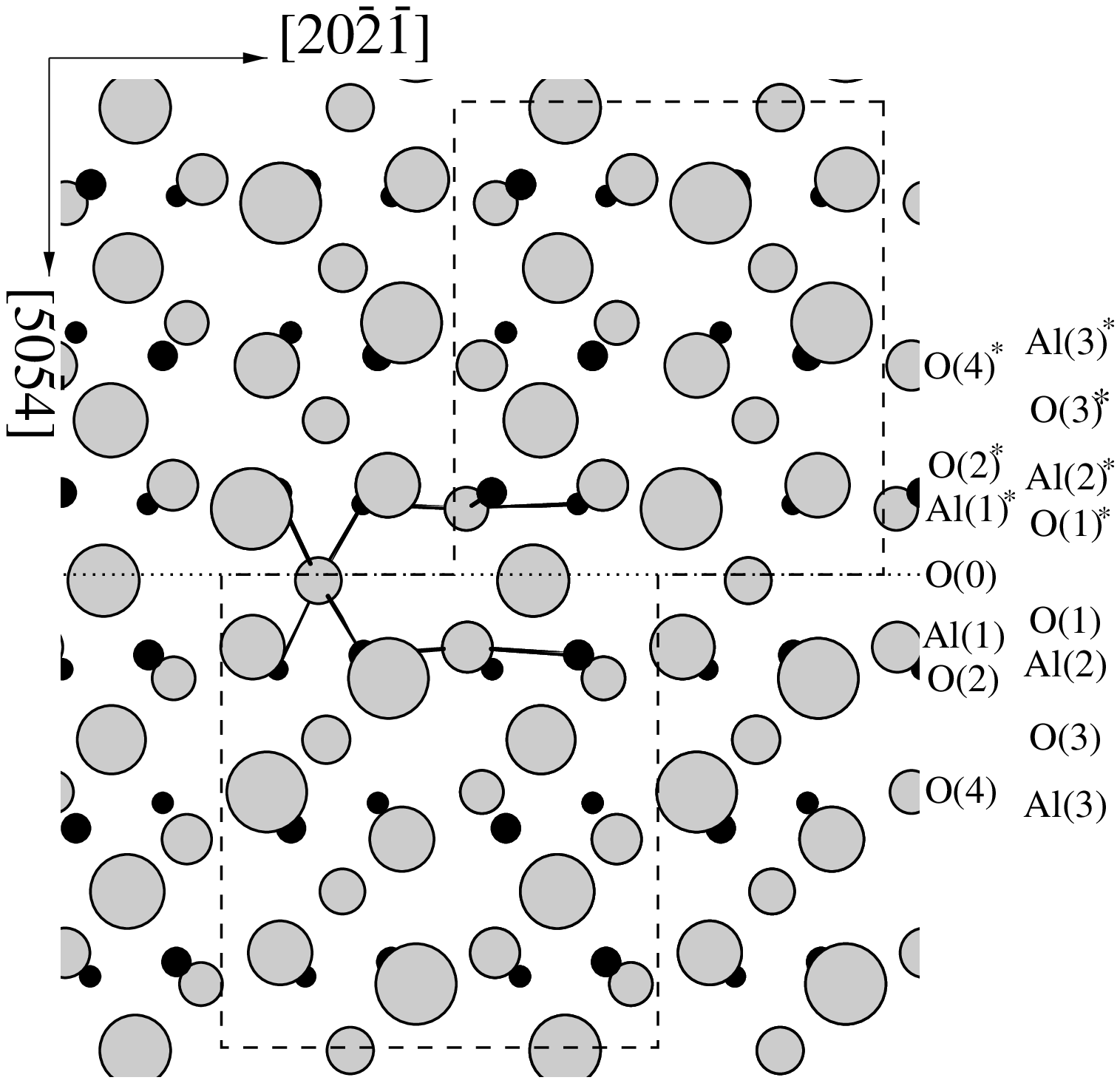,width=8.0cm,angle=0}} 
\end{figure}

 The structure with the second lowest energy is defined by the lateral
 translation state ($T_1,T_2$)=(0.5,0.5). It has an oxygen layer on the
 boundary plane, and the grains are related by a glide-mirror symmetry with
 respect to ($10\bar{1}4$). We will denote it by G(O).  Similarly to the
 Al-terminated case, both the LDFT and shell-model atomic relaxations
 predict the metastability of this translation state: the atomic relaxation
 involves mainly the first three interfacial layers. The relaxed supercell
 is shown in Figure~\ref{afig}. Note the characteristic `V' shape formed by
 the interfacial O atoms (O(1)-O(0)-O(1)$^*$ in Figure~\ref{afig}). A HRTEM
 micrograph of this structure showing the holes as white spots would
 produce a continuous pattern across the boundary.

 Finally, the metastable structure with the highest energy is obtained by
 the translation state ($T_1,T_2$)=(0,0.5), leading to an O-terminated
 configuration with a twofold screw-rotation symmetry around
 [$20\bar{2}\bar{1}$]. It is denoted as S(O).  The relaxed supercell is
 shown in Figure~\ref{SOfig}. Both the LDFT and shell-model atomic relaxations
 show that this translation state is not stable. The atomic forces drive
 the bicrystal towards ($T_1,T_2$)=(0.12,0.5). This is a new metastable
 interface structure that was not noticed in the previous studies of this
 GB. Holes would produce a broken pattern of spots, intermediate between
 the G(0) and G(Al) ones.

 Also for G(O), the adequacy of the axial supercell size was checked by
 comparing the relaxation patterns obtained with 40-, 60-, and 80-atom
 supercells. The LDFT relaxation of the 40-atom cell involved all the atoms
 hence leaving virtually no bulk-like volume remote from the boundary,
 while the shell-model relaxation eliminated the boundaries, reconstructing
 a single-crystal. This procedure indicated that the 40-atom cell is too small
 for the O-terminated interfaces. The LDFT and shell-model relaxations for
 the 60-atom supercell proceeded leaving the innermost atoms in a bulk-like
 environment. Again, the adequacy of the axial supercell size was checked
 by relaxing an 80-atom supercell and by noticing the small difference of 2
 \% between the interface energies of the 60- and 80-atom supercells. The
 final interfacial configuration obtained with the two supercells showed no
 significant difference (smaller than 0.04 \AA\ in all the atomic positions
 within the supercell)

\subsection{Energetics and local atomic structure}

 The supercells for the metastable grain-boundary structures identified in
 the previous section were then fully relaxed with respect to the
 inter-granular separation (excess volume) and to all atomic positions.
 The interface energy was calculated as the difference between the total
 energy of the supercell, and the total energy of an equal number of
 Al$_2$O$_3$ formula units in the bulk phase, divided by the total
 grain-boundary area in the supercell.


\subsubsection{G(Al) interface}
 
 First-principles LDFT calculations for the 60-atom supercell predict a
 small axial excess 
 GB expansion (2\% of $a_{\rm rho}$, where $a_{\rm rho}$=5.086
 \AA\, cf. Ref.~\onlinecite{Marinop00}) of the Al-terminated boundary, which
 lowers the interface energy from 1.92 J/m$^2$ (corresponding to 0\%
 expansion) to 1.88 J/m$^2$ (for 2\% expansion). This interface energy is a
 relatively high value, compared to the ones of the rhombohedral and basal
 twins, namely 0.63 J/m$^2$ (Ref.~\onlinecite{Marinop00}) and 0.73 J/m$^2$
 (Ref.~\onlinecite{Marinop01}).

 The local atomic environment of this interface, which we anticipate is the
 most ordered one among the three metastable structures considered in this
 study, is considerably more disordered than the ones at the rhombohedral
 or basal twins. For example the atomic arrangements on opposite sides of
 the GB are not equivalent (Figure~\ref{cfig}). In terms of Al-O
 coordination, the cations in Al(1) are 5-fold coordinated while the ones
 in Al(1)$^*$ are 6-fold coordinated. The bond lengths are summarized in
 Table~\ref{tabdist} and compared to the ones for bulk alumina.

 In bulk alumina, each Al atom is coordinated by 6 anions at two
 characteristic distances of 1.84 and 1.96 \AA. A similar splitting in the
 interfacial Al-O distances was observed in the most stable of the
 rhombohedral twin structures,~\cite{Marinop01} but not at the present
 $\Sigma 13$ twin interface, where the bond lengths of the shell of first
 neighbours are considerably more spread.

 Similarly, there is a difference in coordination numbers between the
 oxygen atoms in O(2) and O(2)$^*$, which are 4- and 3-fold coordinated
 respectively.  All the other oxygen atoms are 4-fold coordinated. In
 particular, the atomic environment around the innermost O atom is well
 bulk-like with two bond lengths around 1.82 \AA\ and two around 1.95 \AA\
 (cf. bulk values of 1.84 and 1.96 \AA ).

 The effect of atomic relaxations at the interface, compared to the bulk
 volume, is described by the differences in interlayer spacings shown in
 Table~\ref{tabeng}. The relaxation involves mainly the first three atomic
 layers on both sides of the interface. As in the case of the bond lengths,
 there is a small difference between the atomic environment on opposite
 sides of the GB, which is also reflected in the change in interlayer
 distance (Table~\ref{tabeng}). Note again that the central atoms are
 practically unaffected by the relaxations induced by the boundary. These
 results suggest that the boundary effects involve mainly a slab of
 $\approx$ 5 \AA\ about the interface plane.

 The free surface corresponding to this interface, denoted as FS(Al), was
 obtained by artificially forming two grains with Al-terminating
 surfaces. These can be described as obtained by cutting the bicrystal
 structure of Figure~\ref{cfig} between layers Al(1) and Al(1),$^*$ and by
 separating the grains until they do not interact any more. This is
 equivalent to consider only one of the two grains in the supercell and
 repeat it periodically along $T_3$ with a large lattice constant.  The
 structure was then relaxed. The change in interlayer spacing due to
 relaxation is described in Table~\ref{tabeng}. The outermost Al layer
 contracts inward, while the outermost O layer relaxes outward, so that the
 final surface configuration is O-terminated. The corresponding calculated
 surface energy is 2.77 J/m$^2$, a relatively high value compared to
 the ones of the basal surface (1.94 J/m$^2$ obtained by us with the present
 approach, which was also used for the basal twins in
 Ref.\onlinecite{Marinop01}; 1.95 J/m$^2$ Ref.\onlinecite{Batirev99prl};
 2.13 J/m$^2$ Ref.\onlinecite{Wang00}) and the rhombohedral surface
 (1.98 J/m$^2$ calculated with the present approach in 
 Ref.\onlinecite{Marinop00}).

\subsubsection{G(O) interface}

 The axial relaxation of the G(O) interface, calculated by LDFT, yields
 that the minimum energy of 2.44 J/m$^2$ corresponds to an axial excess GB
 expansion of 6\% of $a_{\rm rho}$. The change in interlayer
 spacing due to atomic relaxation for the 60-atom cell is shown in
 Table~\ref{tabeng}. It suggests that the relaxation involves a slab of
 $\approx$ 7 \AA\ centered at the interface plane. In this case, the atomic
 environments on opposite sides of the boundary are equivalent. The
 boundary atoms O(0) are the most affected by the relaxation procedure, but
 also the inner O(2) and O(3) relax quite substantially (displacements of
 4.6 \% of $a_{\rm rho}$).


 The local atomic environment is summarized in Table~\ref{tabdist}. Despite
 the large atomic relaxations, the interfacial oxygen O(0) are in a
 bulk-like environment, being 4-coordinated with two bond-lengths at 1.84
 \AA\ and two at 1.99 \AA\ (cf. 1.84 \AA\ and 1.96 \AA\ in bulk). On the
 contrary, the next layers O(1) and O(2), and the inner layer O(4) are
 3-fold coordinated. (The fourth bond from O(4), which points towards 
 the interface, is elongated to
 2.11 \AA . This is the case also for the 80-atom cell. Hence, the
 60-atom cell is appropriate.)  Similarly, the
 first three cations layers Al(1), Al(2), and Al(3) are under-coordinated
 with only 5 oxygen neighbors. In the 80-atom supercell, the first atoms in
 bulk-like environment are the ones in Al(4), which are 6-fold coordinated,
 with three bond-lengths centered at 1.85 \AA\ and three at 1.94 \AA.

 There are no obvious ways of defining a free surface from the interface
 configuration shown in Figure~\ref{afig}. One possible variant, that has
 the advantage of creating stoichiometric and equivalent surfaces, is to
 assign half of the interfacial atoms O(0) to one grain and half to the
 others. This creates a starting anion terminated configuration with a
 jagged structure. The structural minimization of this free surface leads
 to an oxygen terminated structure, indicated as FS(O), with a surface
 energy of 2.50 J/m$^2$, hence being more stable than FS(Al). The changes
 in interlayer spacing due to relaxation are included in
 Table~\ref{tabeng}.
  
\subsubsection{S(O) interface}

\begin{figure}
\caption{LDFT result for the relaxed atomic structure of the S(O)
interface viewed along the [$1\bar{2}10$] direction. Atoms, labels, and
symbols as in Figure~\ref{cfig}.}
\label{SOfig}
 \centerline{\psfig{file=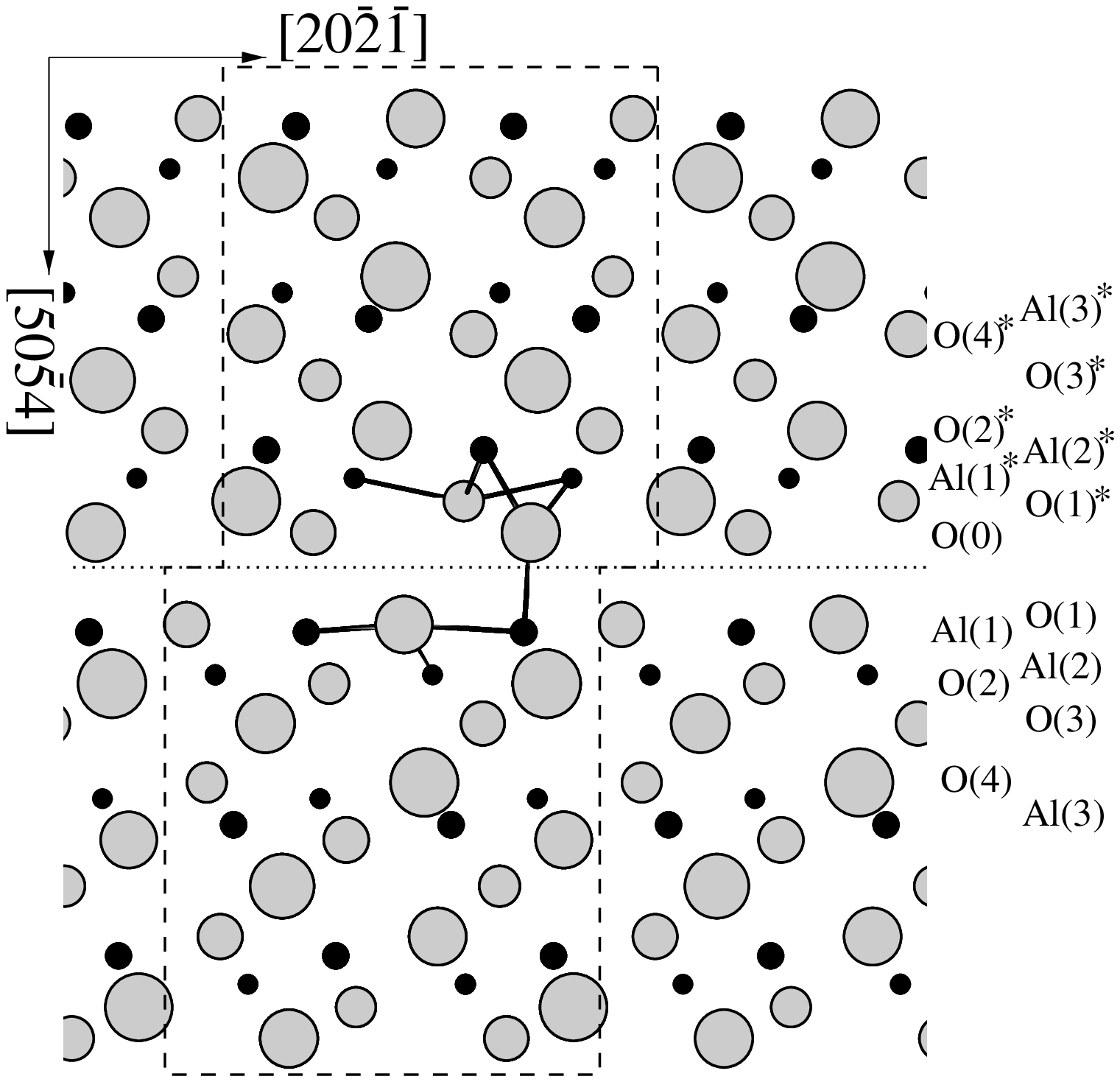,width=8cm,angle=0}} 
\end{figure}

 We already discussed that the initially imposed translation state
 $(0,0.5)$ for this interface is not stable. Let us now specify this
 statement more precisely: the starting configuration $(0,0.5)$
 identified with the shell-model calculations brings the oxygen atoms O(0)
 and O(1) very close together. We therefore started the LDFT structural
 optimization with a larger inter-granular distance $T_3$ (6 \% of $a_{\rm
 rho}$). The atomic relaxation increased the distance between the ``close''
 O(0) and O(1) layers, and this induced also a change in the lateral
 translation state. The minimum energy of 2.71 J/m$^2$ is obtained at a
 rather large inter-granular separation, 12 \% of $a_{\rm rho}$, which
 produces a characteristic open boundary structure shown in
 Figure~\ref{SOfig}. In particular, there are large open channels along the
 boundary in the [$20\bar{2}\bar{1}$] direction, with a typical diameter of
 6.7 \AA. Here impurity or dopant atoms of different species could in
 principle be accommodated interstitially without involving severe elastic
 strains.

 Atomic relaxations involve mainly the interfacial oxygen layer O(0) and
 the first three atomic layers on both sides. The changes in interlayer
 spacings are summarized in Table~\ref{tabeng}. These results suggest that
 the boundary affects mainly a slab of $\approx$ 6 \AA\ centered at the
 interface.

 The innermost atoms Al(3) are practically unaffected by the interfacial
 relaxations, and thus they are in a bulk-like environment. The atomic
 distances of the first neighbouring O shell have the characteristic
 bulk-like splitting, three bond lengths being around 1.85 \AA\ and three
 around 1.94 {\AA}. On the interface, cations and anions are respectively
 5-fold and 3-fold coordinated. However, the anions of the second layer
 O(2) have already the bulk coordination number of 4. The Al-O bond lengths
 are summarized in Table~\ref{tabdist}.

 The difference in interlayer separation between O(0)-O(1) and
 O(0)-O(1)$^*$, respectively 1.6 and 0.78 \AA (+8.5 \% and -9.8 \% of
 a$_{\rm rho}$, respectively, cf. Table \ref{tabdist}), suggests the
 O(0)-O(1) plane as an alternative cleavage plane to the ones already
 mentioned. The two free surfaces obtained by cleaving the bicrystal of
 Figure~\ref{SOfig} at these planes are not stoichiometric and
 geometrically not equivalent, but the supercell is overall
 stoichiometric. Both these FS contribute to an average surface energy of
 2.76 J/m$^2$. This average energy value provides some information about
 the possible cleavage mechanisms of alumina bicrystals, which we will
 discuss in the next section. 
 Incidentally its value is close to the one of the surface energy of the
stoichiometric Al-terminated surface (see section IV.B.1).

\subsection{Large inter-granular separations: bonding and cleavage}

 Two sets of LDFT calculations monitored the interface energy as a function
 of large inter-granular separation in two different regimes. The starting
 points of the first one were the isolated grains with free surfaces, which
 were then brought closer, down to the formation of the three metastable
 interfaces previously described. These calculations are conceptually
 similar to the experimental procedure of diffusion bonding of bicrystals
 in ultrahigh vacuum, and should shed light on the boundary formation
 mechanisms. The starting points of the second set of calculations were the
 relaxed metastable translation states of the bicrystals, described in the
 previous section, which were then gradually pulled apart, up to the
 formation of the free surfaces. These calculations reveal insights
 into the cleavage mechanisms.
 
 The fundamental difference between the two methods is that in the first
 case the cleavage plane is defined by an {\it a priori} choice of the
 free-surface terminating layer, while in the second case the cleavage
 plane is determined by the mechanical properties of the bicrystal, and
 therefore by the weakest bonds.
 
\begin{figure}
\caption{Interface energies (in J/m$^2$) versus the axial intergranular
separation (in \% of $a_{\rm rho}$) for the three metastable interfaces
G(Al), G(O), and S(O). The solid (dashed) lines connect the results
obtained by reducing (increasing) the intergranular separation from free
surfaces (internal interfaces) down to the the formation of the internal
boundaries (free surfaces).  Solid horizontal lines represent twice the
energies of the free surfaces FS(Al) and FS(O) (cf. Table~\ref{tabeng}; 
NB: upon interfacial separation the interface energy 
goes to twice the surface energy because the resulting 
surface area is twice the initial interface area). 
}
\label{envol}
 \centerline{\psfig{file=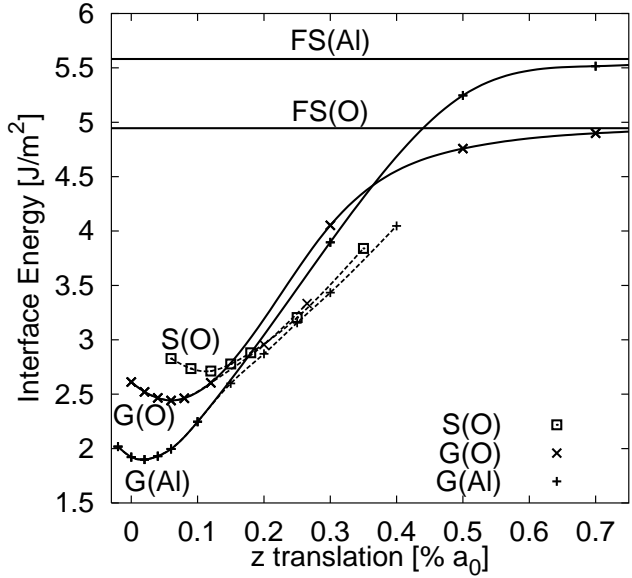,width=8.5cm,angle=-90}} 
\end{figure}

 The LDFT results for the interface energies versus the axial component
 $T_3$ are shown in Figure~\ref{envol}. The energies corresponding to the
 first set of calculations (from free surfaces to internal boundaries) are
 connected with solid lines, while the values for the second set (from
 internal boundaries to free surfaces) are connected with dashed
 lines. Solid horizontal lines represent the limit of free surfaces, being
 twice the energy of the free stoichiometric surfaces FS(Al) and FS(O),
 5.53 and 4.94 J/m$^2$ respectively. Each energy value corresponds to a
 fully relaxed configuration of atoms in the supercell.

 Let us first concentrate on intergranular bonding, i.e., the formation of
 the two most stable structures, G(Al) and G(O), from the corresponding
 free stoichiometric surfaces FS(Al) and FS(O), by reducing the
 intergranular distance. Note that the relative energetic stability of the
 free surfaces is reversed with respect to that one of the bicrystals:
 FS(O) is more stable than FS(Al), and G(Al) is more stable than G(O). With
 decreasing distance, the interface energies remain close to the values of
 the corresponding free surfaces down to an intergranular distance of 2.54
 \AA\ (0.5 \% of a$_{\rm rho}$). There a significant difference of more
 than 5 \% appears due to the onset of interaction between the two slabs,
 and from there the values monotonically decrease down to the equilibrium
 metastable configurations G(Al) and G(O).  The crossover point in the
 relative energetic stability, when the Al termination becomes more stable
 than the O termination, is around 1.8 \AA\ (0.35 \% of a$_{\rm rho}$).

 For intergranular cleavage, starting from the equilibrium interfaces G(Al)
 and G(O), and stretching the bicrystals along $T_3$, the reverse energy
 paths as for the bonding procedure are followed up to displacements of
 0.81 \AA\ (0.16 \% of a$_{\rm rho}$). Subsequently, both the bicrystals
 react to the increase in intergranular distance by stretching the weakest
 bonds: the cleavage plane passes through O(1)$^*$-O(2)$^*$ in G(Al) and
 through O(0)-O(1) in G(O). Cleavage on these planes create
 non-stoichiometric free surfaces, leaving two terminating oxygen atoms on
 one side and one on the other side. 

At larger intergranular distances, the energies of the three metastable
structures increase at the same rate, apparently to the same
surface-energy limit. This may be understood by noticing that, in the
limit of large separation, the two grains resulting from
this `asymmetric' cleavage in the G(Al) and G(O) supercells are
structurally equal to the grains resulting from the cleavage 
through O(0)-O(1) in the S(O) supercell. Therefore, the corresponding three
cleaved free surfaces become the same for large separation 
in the supercells. Their surface energies become equal to 2.26 $J/m^2$,
which is the average surface energy of the two non-stoichiometric terminating
surfaces identified for S(O) (see section IV.B.3).

\subsection{Electronic structure}

 The self-consistent LDFT results obtained for the three equilibrium
 metastable interface configurations were used to investigate the effect of
 the interfacial local atomic arrangements on the electronic structures of
 alumina bicrystals. In particular, we focus on the site-projected density
 of states (PDOS). We define the PDOS as the projection of the total
 density of states (DOS) on partial waves inside spheres centered on atomic
 sites. The radius of the spheres were 1.27 \AA\ and 0.58 \AA\ for oxygen
 and aluminium sites, respectively. The crystal Bloch eigenstates were
 calculated on a $6 \times 6 \times 2$ {\bf k}-point mesh and broadened
 with Gaussians of 0.8 eV width. The same setup was used in the analysis of
 the rhombohedral $\Sigma 7$ and basal $\Sigma 3$ twin
 boundaries.~\cite{Marinop00,Marinop01}

\begin{figure}
\caption{a) O-PDOS for the G(Al), b) for the G(O), and c) for the S(O)
interfaces. The zero of energy is the Fermi level (energy eigenvalue of the
highest occupied valence-band state) of the bulk supercell, the
interfaces PDOS have been rigidly shifted so that the bulk-like O-PDOS
(thin solid lines) is on top of the O-PDOS for bulk alumina (thick solid
line). The Fermi levels of the supercells containing the interfaces are
marked by solid vertical lines.}
\label{dos}
 \centerline{\psfig{file=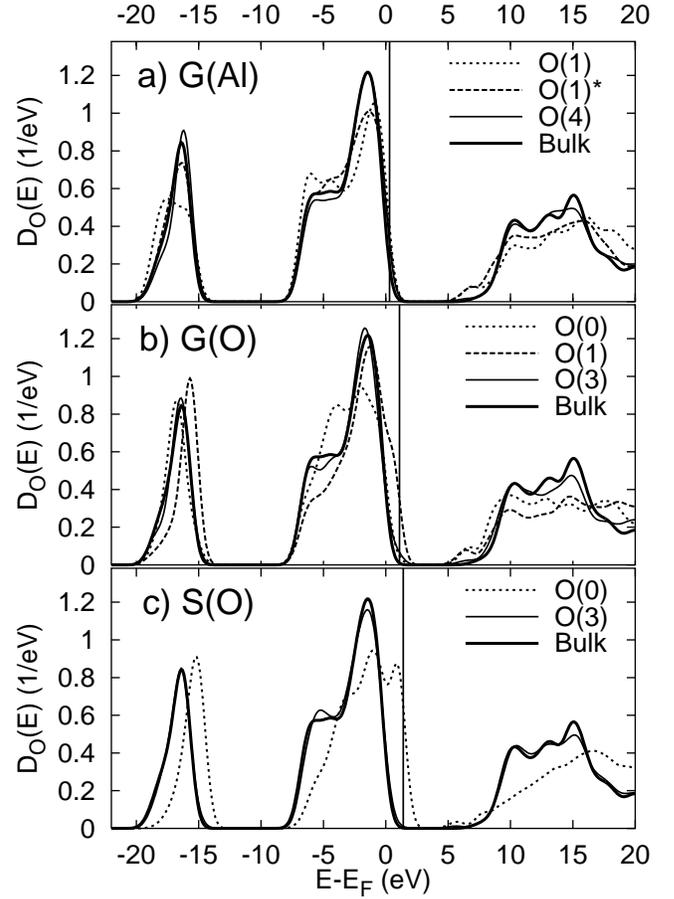,width=9cm,angle=0}} 
\end{figure}

 The results are shown in Figure~\ref{dos}, where the oxygen-PDOS (O-PDOS)
 of the three interface states are compared with the O-PDOS of bulk
 alumina. The latter is indicated by a thick solid line, while a dashed
 line and a thin solid line indicate the PDOS of oxygen atoms, close and
 far from the interface, respectively. To facilitate the discussion of the
 results, the interfaces PDOS have been rigidly shifted so that the
 bulk-like O-PDOS (thin solid lines) is on top of the O-PDOS for bulk
 alumina (thick solid line). The zero of energy in the plot is the Fermi
 level of the bulk supercell (the energy eigenvalue of the highest occupied
 valence state), and the Fermi levels of the supercells containing the
 interfaces are marked with vertical solid lines.


 A discussion of the Al-PDOS is not given here because, as already
 considered in Ref.~\onlinecite{Marinop00}, it is less illustrative than
 the O-PDOS in \aal2o3: Due to the high ionic character of the Al-O bond in
 alumina, all the features of the DOS (up to $\approx$ 20 eV) are dominated
 by the O-PDOS, the contribution of the Al-PDOS being 5$-$10 times smaller,
 and similar in the spectral shape of peaks and shoulders, even in the
 lower conduction band region.

 Let us first emphasize the similarity between the O-PDOS in bulk alumina
 (thick solid lines) with the ones in bulk-like environment far from the
 interfaces (thin solid lines). There are hardly any differences in the
 lower valence band (with energies centered around -17 eV), and minor
 differences in the higher valence band (with energies centered around -5
 eV) and in the conduction band. This shows that the effect of the local
 environment on the electronic structure is localized within the first 2-3
 interfacial oxygen layers, the most affected being always the interfacial
 ions (dashed lines), as intuitively expected and already observed at the
 rhombohedral $\Sigma 7$ twin boundary in Ref.~\onlinecite{Marinop00}.
 This further confirms that the supercell dimension along $T_3$ is
 sufficiently large to leave the inner atoms in a bulk-like environment.

 We now discuss the differences between the interfacial and bulk-like
 valence O-PDOS. Let us first notice that the $\Sigma 13$ GB affect both
 the lower and higher valence bands of the interfacial O atoms. The
 differences are moderate in G(Al), where the interfacial O-PDOS
 corresponding to O(1) and O(1)$^*$ (dashed lines in Figure~\ref{dos}a),
 have still the characteristic bulk-like structure (thin solid line). The
 biggest deviation appears for O(1),$^*$ which is 5-fold coordinated. More
 pronounced differences are present in G(O), where the interfacial O-PDOS
 (O(0) and O(1)=O(1)$^*$, dashed lines in Figure~\ref{dos}b) are
 qualitatively different from the bulk-like structure (thin solid line):
 the states corresponding to O(1) (3-fold coordinated) are redistributed
 towards higher energies, while the states corresponding to the O(0)
 (4-fold coordinated) higher valence bands are redistributed towards lower
 energies, forming a two-peak structure.  The high-energy S(O) boundary
 induces considerably more drastic changes in the electronic structure. The
 lower valence band is rigidly shifted by 1.8 eV to higher energies, and a
 three-peak spectral shape appears in the higher band.  Let us also note
 that the valence-band edges for the G(Al), G(O), and S(O) supercells,
 respectively, are 0.2 eV, 0.9 eV, and 1.3 eV higher than for bulk alumina.

%

 The changes in the unoccupied region of the O-PDOS are now addressed. The
 main feature, common to the three interfaces, is the presence of a
 pre-peak in the energy gap, at 6-7 eV, just below the conduction-band
 edge. A similar gap state was observed in the highest energy structure of
 the rhombohedral twin.  The fundamental band gap of bulk alumina (6.6 eV
 with the LDFT method used in the present calculations) is therefore
 reduced by the presence of this pre-peak to 5.41 eV in G(Al), 4.47 eV in
 G(O) and 3.72 eV in S(O).

\begin{figure}
\caption{Real-space projection along the [$1\bar{2}10$] direction 
of the density distribution of unoccupied electron
 states in the energy range between 5 and 7 eV for the G(Al)
 interface. Al and O labels mark the projected 
atomic positions in the supercell.}
\label{gstate}
 \centerline{\psfig{file=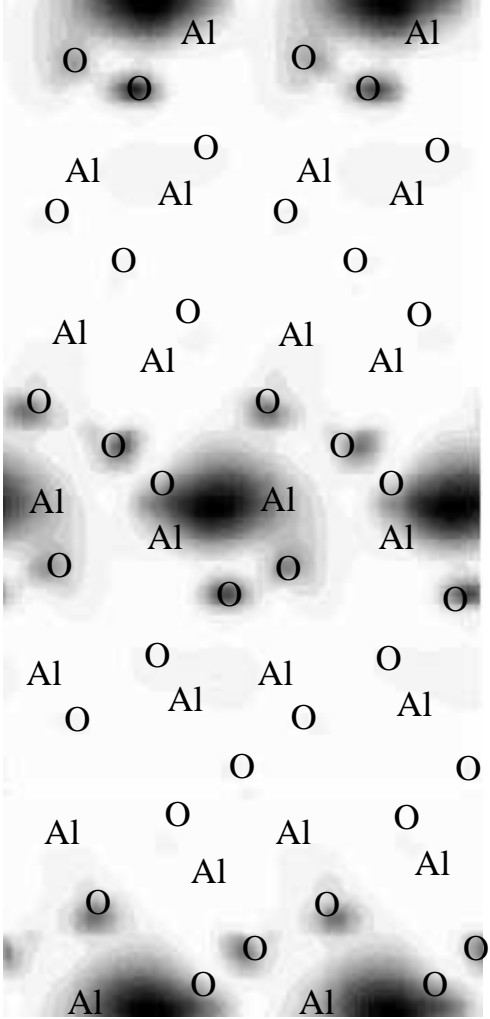,width=6.5cm,angle=0}} 
\end{figure}

 The real-space projection along [$1\bar{2}10$] of the density of
 unoccupied electron states in the energy range between 5 and 7 eV is shown
 in Figure~\ref{gstate} for the G(Al) interface, taken as the
 representative case. The density is proportional to the grey-scale,
 ranging from white for zero density, to black for the highest density.  The
 projected atomic positions in the supercell are marked by the labels Al
 and O. Note that most density of the electron states corresponding to the
 pre-peak is spatially confined at the interface, and that a considerable
 density fraction is localized on the first three interfacial O atoms.

 At higher energies, between 10 and 20 eV, the three interfaces modify the
 interfacial O-PDOS substantially with respect to the bulk three-peak
 signal. Moreover, within the same boundary structure, there are
 significant differences between the O-PDOS corresponding to different
 interfacial sites, like O(1) and O(1)$^*$ in G(Al), or O(0) and O(1) in
 G(O). This is likely to play an important role in the interpretation of
 experimental ELNES spectra (cf, e.g. Ref~\onlinecite{Nufer01}).
 

 As demonstrated in Ref.~\onlinecite{Nufer01} for the rhombohedral $\Sigma
 7$ twin boundary, the ELNES of the oxygen core-level ionisation $K$ edge,
 which is approximately proportional to the conduction band PDOS of
 angular-momentum p character at oxygen sites, can provide experimental
 informations to distinguish the local electronic states and atomic
 coordination at oxygen sites located in the bulk volume or at
 interfaces. The calculated O-PDOS for the unoccupied conduction-band
 states of the $\Sigma 13$ twin boundary indicate characteristic
 differences between the three metastable structural models obtained by the
 LDFT structure optimisation. Hence, an ELNES experiment with a $\Sigma 13$
 twin bicrystal in a scanning transmission electron microscope with
 sufficiently high spatial and energetic resolutions, comparably to the
 ELNES experiment of Ref.~\onlinecite{Nufer01}, appears very promising to
 augment our theoretical characterisation of this interface.

\section{Conclusions}
\label{concl}

 The energetic, structural, and electronic properties of the $\Sigma 13$
 ($10\bar{1}4$) mirror twin boundary in \aal2o3 were studied by atomistic
 modelling with both empirical shell-model potentials and first-principles
 LDFT calculations. Three metastable variants of the interface were
 identified, one Al-terminated boundary, G(Al), and two O-terminated
 boundaries, G(O) and S(O). The lowest-energy interface is Al-terminated,
 as previously predicted by empirical shell-model
 potentials.~\cite{Kenway94,Hoche94a} The interfacial energies of the fully
 relaxed interfaces, as calculated with LDFT, were discussed.


 Despite the thermodynamic stability of the G(Al) interface over the
 O-terminating variants, a high-resolution transmission electron microscopy
 (HRTEM) experiment with high-purity diffusion-bonded bicrystals detected
 the G(O) boundary, rather then the G(Al) one.~\cite{Hoche94a} In order to
 clarify these observations, the formation of the G(Al) and G(O) boundaries
 was studied by gradually reducing the intergranular distance, starting
 from the corresponding free surfaces FS(Al) and FS(O), to model
 theoretically the procedure of diffusive bonding in ultrahigh vacuum. This
 showed that the relative energetic stability between these two interfaces
 depends on the intergranular separation, and is actually reversed for the
 free surfaces, FS(O) being more stable than FS(Al). From this a possible
 explanation for the HRTEM observations emerges: prior to bonding the free
 surfaces are O-terminated. When the crystals are diffusion bonded, the
 equilibrium thermodynamics would drive the atomic reconstruction of the
 O-terminated interface into the Al-terminated one, G(Al). However, the
 energy barrier for such a reconstructive transformation is apparently too
 high, as long as no additional Al is provided in the bonding process, and
 the bicrystal is therefore locked in the metastable G(O) configuration.

 The analysis of the site-projected densities of states showed that the
 effect of the interfaces on the electronic structure is localized within
 5-7 \AA\ around the boundary plane, and affects mostly the upper valence
 band and the conduction bands.  In particular, the three variants of the
 $\Sigma 13$ boundary narrow the band gap by defect states below the
 conduction-band edge, which are highly localized at the interface.
 
 An important final conclusion of this work is that, although the $\Sigma
 13$ twin boundary has a short lateral periodicity (which enables the
 thorough treatment by LDFT), the resulting structural complexities, as
 well as the interface energies of its metastable configurations, are
 significantly higher than those of the previously investigated
 rhombohedral $\Sigma 7$ and basal $\Sigma 3$ twin boundaries, usually
 called ``special'' grain boundaries.  The $\Sigma 13$ interface is
 therefore a promising model case for ``general'' grain boundaries in
 \aal2o3. We anticipate that it will be useful to provide a theoretical
 understanding of the influence of segregated cation impurities on the
 structure and energetics of alumina interfaces.


 
\acknowledgements

 This project was financially  
 supported by the Deutsche Forschungsgemeinschaft (project
 El 155/4-1). The authors thank A. G. Marinopoulos for his help in the
 beginning of the work, S. Nufer for valuable communications concerning
 ELNES experiment and theory, T. H\"{o}che, M. W. Finnis and 
 M. R\"{u}hle for helpful discussions. 


\begin{table}
\caption{Equilibrium Al-O distances in the interfacial and bulk-like
regions of the G(Al), G(O), and S(O) supercells. For comparison, there
are two Al-O distances in bulk \aal2o3, with calculated
values of 1.84 {\AA} and 1.96 {\AA}.}
\label{tabdist}
\begin{center}
 \begin{tabular}{cccc}
\multicolumn{3}{c}{GB region}    & bulk-like region \\ \hline
\multicolumn{4}{c}{\it G(Al)}                           \\ 
 Al(1)$^*$-O     &      & Al(1)-O         & O(5)-Al           \\
1.82 1.90 1.93   &      & 1.81 1.82      &  1.80 1.85        \\
1.94 1.96 2.00   &      & 1.93 1.95 2.00 &  1.94 1.97        \\ \hline
\multicolumn{4}{c}{\it G(O)}                            \\  
 O(1)$^*$-Al     & O(0)-Al   & O(1)-Al   & Al(3)-O        \\
1.77 1.80        & 1.84 (2x) & 1.77 1.80 & 1.83 1.84 1.87 \\
1.92             & 1.99 (2x) & 1.92      & 1.93 1.95 1.97 \\ \hline
\multicolumn{4}{c}{\it S(O)}                            \\  
O(1)$^*$-Al      &  O(0)-Al  & O(1)-Al   & Al(3)-O             \\
1.81 1.88        & 1.77 1.77 & 1.76 1.81 & 1.84 1.84 1.89             \\
1.93             & 1.81      & 2.01      & 1.93 1.94 1.97             \\
  \end{tabular}
\end{center}
\end{table}

\begin{table*}
\caption{LDFT results for interface energies, axial expansions along $T_3$
 (in \% of a$_{\rm rho}$) and changes in interlayer spacings (in \% of
 a$_{\rm rho}$) for the structurally relaxed twin interfaces and free
 surfaces.}
\label{tabeng}
\begin{center}
 \begin{tabular}{lcc|lccc}
                         & G(Al) &  FS(Al) & 
                         & G(O)  & S(O) & FS(O) \\ \hline
$E_{\rm int}$  (J/m$^2$) & 1.88  & 2.77    & 
$E_{\rm int}$  (J/m$^2$) & 2.44  & 2.71 &  2.47 \\
$T_3$                    & 2     & -       & 
$T_3$                    & 6     & 12  & - \\ \hline

Al(3)$^*$-O(4)$^*$       & 0.2   &  0.5    &
O(4)$^*$-Al(3)$^*$       & -0.5  & -0.5    & -1.5 \\

Al(2)$^*$-Al(3)$^*$      & -0.3  &  0.7    &
O(3)$^*$-O(4)$^*$        &  0.6  & -0.3    & 1.5  \\

O(3)$^*$-Al(2)$^*$       & -1.1  &  1.9    &
O(2)$^*$-O(3)$^*$        &  4.6  &  0.6    & 0.9  \\

O(2)$^*$-O(3)$^*$        & -2.5  & -3.0    &
Al(2)$^*$-O(2)$^*$       & -2.2  &  1.1    & -3.8  \\

O(1)$^*$-O(2)$^*$        & 2.1   &  10.1   & 
Al(1)$^*$-Al(2)$^*$      & -3.9  & -0.4    & -0.8 \\

Al(1)$^*$-O(1)$^*$       & 0     & -11.0   & 
O(1)$^*$-Al(1)$^*$       & -2.8  &  1.3    & 3.8 \\

                         &       &          & 
O(0)-O(1)$^*$            & 7.3   & -9.8    & 0.6 \\

Al(1)$^*$-Al(1)          & 1.7   & $\infty$ & 
                         &       &          & \\
 
                         &       &          & 
O(0)-O(1)                & 7.3   &  8.5     & 0.6 \\

Al(1)-O(1)               & 3.0   & -11.0   & 
O(1)-Al(1)               & -2.8  & -3      & 3.8 \\

O(1)-O(2)                & -6.1  &  10.1   & 
Al(1)-Al(2)              & -3.9  &  5.6    & -0.8 \\

O(2)-O(3)                &  3.0  & -3.0    &
Al(2)-O(2)               & -2.2  & -2.9    & -3.8 \\

O(3)-Al(2)               &  0.4  &  1.9    &
O(2)-O(3)                &  4.6  & -2.5    &  0.9 \\

Al(2)-Al(3)              & -1.4  &  0.7    &
O(3)-O(4)                &  0.6  &  3.5    &  1.5 \\

Al(3)-O(4)               & 0.6   & 0.5     &
O(4)-Al(3)               & -0.5  & -0.1    & -1.5 

  \end{tabular}
\end{center}
\end{table*}

\newpage

\newpage

\newpage

\newpage

\newpage

\end{document}

%% file: psfig2
\def\PsfigVersion{1.9}
\ifx\undefined\psfig\else \fi

\let\LaTeXAtSign=\@
\let\@=\relax
\edef\psfigRestoreAt{\catcode`\@=\number\catcode`@\relax}
\catcode`\@=11\relax
\newwrite\@unused
\def\ps@typeout#1{{\let\protect\string\immediate\write\@unused{#1}}}
\ps@typeout{psfig/tex \PsfigVersion}

\def\figurepath{./}

\def\@nnil{\@nil}
\def\@empty{}
\def\@psdonoop#1\@@#2#3{}
\def\@psdo#1:=#2\do#3{\edef\@psdotmp{#2}\ifx\@psdotmp\@empty \else
    \expandafter\@psdoloop#2,\@nil,\@nil\@@#1{#3}\fi}
\def\@psdoloop#1,#2,#3\@@#4#5{\def#4{#1}\ifx #4\@nnil \else
       #5\def#4{#2}\ifx #4\@nnil \else#5\@ipsdoloop #3\@@#4{#5}\fi\fi}
\def\@ipsdoloop#1,#2\@@#3#4{\def#3{#1}\ifx #3\@nnil 
       \let\@nextwhile=\@psdonoop \else
      #4\relax\let\@nextwhile=\@ipsdoloop\fi\@nextwhile#2\@@#3{#4}}
\def\@tpsdo#1:=#2\do#3{\xdef\@psdotmp{#2}\ifx\@psdotmp\@empty \else
    \@tpsdoloop#2\@nil\@nil\@@#1{#3}\fi}
\def\@tpsdoloop#1#2\@@#3#4{\def#3{#1}\ifx #3\@nnil 
       \let\@nextwhile=\@psdonoop \else
      #4\relax\let\@nextwhile=\@tpsdoloop\fi\@nextwhile#2\@@#3{#4}}
\ifx\undefined\fbox
\newdimen\fboxrule
\newdimen\fboxsep
\newdimen\ps@tempdima
\newbox\ps@tempboxa
\long\def\fbox#1{\leavevmode\setbox\ps@tempboxa\hbox{#1}\ps@tempdima\fboxrule
    \advance\ps@tempdima \fboxsep \advance\ps@tempdima \dp\ps@tempboxa
   \hbox{\lower \ps@tempdima\hbox
  {\vbox{\hrule height \fboxrule
          \hbox{\vrule width \fboxrule \hskip\fboxsep
          \vbox{\vskip\fboxsep \box\ps@tempboxa\vskip\fboxsep}\hskip 
                 \fboxsep\vrule width \fboxrule}
                 \hrule height \fboxrule}}}}
\fi
\newread\ps@stream
\newif\ifnot@eof       
\newif\if@noisy        
\newif\if@atend        
\newif\if@psfile       
{\catcode`\%=12\global\gdef\epsf@start{
\def\epsf@PS{PS}
\def\epsf@getbb#1{%
\openin\ps@stream=#1
\ifeof\ps@stream\ps@typeout{Error, File #1 not found}\else
   {\not@eoftrue \chardef\other=12
    \def\do##1{\catcode`##1=\other}\dospecials \catcode`\ =10
    \loop
       \if@psfile
	  \read\ps@stream to \epsf@fileline
       \else{
	  \obeyspaces
          \read\ps@stream to \epsf@tmp\global\let\epsf@fileline\epsf@tmp}
       \fi
       \ifeof\ps@stream\not@eoffalse\else
       \if@psfile\else
       \expandafter\epsf@test\epsf@fileline:. \\%
       \fi
          \expandafter\epsf@aux\epsf@fileline:. \\%
       \fi
   \ifnot@eof\repeat
   }\closein\ps@stream\fi}%
\long\def\epsf@test#1#2#3:#4\\{\def\epsf@testit{#1#2}
			\ifx\epsf@testit\epsf@start\else
\ps@typeout{Warning! File does not start with `\epsf@start'.  It may not be a PostScript file.}
			\fi
			\@psfiletrue} 
{\catcode`\%=12\global\let\epsf@percent=
\long\def\epsf@aux#1#2:#3\\{\ifx#1\epsf@percent
   \def\epsf@testit{#2}\ifx\epsf@testit\epsf@bblit
	\@atendfalse
        \epsf@atend #3 . \\%
	\if@atend	
	   \if@verbose{
		\ps@typeout{psfig: found `(atend)'; continuing search}
	   }\fi
        \else
        \epsf@grab #3 . . . \\%
        \not@eoffalse
        \global\no@bbfalse
        \fi
   \fi\fi}%
\def\epsf@grab #1 #2 #3 #4 #5\\{%
   \global\def\epsf@llx{#1}\ifx\epsf@llx\empty
      \epsf@grab #2 #3 #4 #5 .\\\else
   \global\def\epsf@lly{#2}%
   \global\def\epsf@urx{#3}\global\def\epsf@ury{#4}\fi}%
\def\epsf@atendlit{(atend)} 
\def\epsf@atend #1 #2 #3\\{%
   \def\epsf@tmp{#1}\ifx\epsf@tmp\empty
      \epsf@atend #2 #3 .\\\else
   \ifx\epsf@tmp\epsf@atendlit\@atendtrue\fi\fi}

\chardef\psletter = 11 
\chardef\other = 12

\newif \ifdebug 
\newif\ifc@mpute 
\c@mputetrue 

\let\then = \relax
\def\r@dian{pt }
\let\r@dians = \r@dian
\let\dimensionless@nit = \r@dian
\let\dimensionless@nits = \dimensionless@nit
\def\internal@nit{sp }
\let\internal@nits = \internal@nit
\newif\ifstillc@nverging
\def \Mess@ge #1{\ifdebug \then \message {#1} \fi}

{ 
	\catcode `\@ = \psletter
	\gdef \nodimen {\expandafter \n@dimen \the \dimen}
	\gdef \term #1 #2 #3%
	       {\edef \t@ {\the #1}
		\edef \t@@ {\expandafter \n@dimen \the #2\r@dian}%
		\t@rm {\t@} {\t@@} {#3}%
	       }
	\gdef \t@rm #1 #2 #3%
	       {{%
		\count 0 = 0
		\dimen 0 = 1 \dimensionless@nit
		\dimen 2 = #2\relax
		\Mess@ge {Calculating term #1 of \nodimen 2}%
		\loop
		\ifnum	\count 0 < #1
		\then	\advance \count 0 by 1
			\Mess@ge {Iteration \the \count 0 \space}%
			\Multiply \dimen 0 by {\dimen 2}%
			\Mess@ge {After multiplication, term = \nodimen 0}%
			\Divide \dimen 0 by {\count 0}%
			\Mess@ge {After division, term = \nodimen 0}%
		\repeat
		\Mess@ge {Final value for term #1 of 
				\nodimen 2 \space is \nodimen 0}%
		\xdef \Term {#3 = \nodimen 0 \r@dians}%
		\aftergroup \Term
	       }}
	\catcode `\p = \other
	\catcode `\t = \other
	\gdef \n@dimen #1pt{#1} 
}

\def \Divide #1by #2{\divide #1 by #2} 

\def \Multiply #1by #2
       {{
	\count 0 = #1\relax
	\count 2 = #2\relax
	\count 4 = 65536
	\Mess@ge {Before scaling, count 0 = \the \count 0 \space and
			count 2 = \the \count 2}%
	\ifnum	\count 0 > 32767 
	\then	\divide \count 0 by 4
		\divide \count 4 by 4
	\else	\ifnum	\count 0 < -32767
		\then	\divide \count 0 by 4
			\divide \count 4 by 4
		\else
		\fi
	\fi
	\ifnum	\count 2 > 32767 
	\then	\divide \count 2 by 4
		\divide \count 4 by 4
	\else	\ifnum	\count 2 < -32767
		\then	\divide \count 2 by 4
			\divide \count 4 by 4
		\else
		\fi
	\fi
	\multiply \count 0 by \count 2
	\divide \count 0 by \count 4
	\xdef \product {#1 = \the \count 0 \internal@nits}%
	\aftergroup \product
       }}

\def\r@duce{\ifdim\dimen0 > 90\r@dian \then   
		\multiply\dimen0 by -1
		\advance\dimen0 by 180\r@dian
		\r@duce
	    \else \ifdim\dimen0 < -90\r@dian \then  
		\advance\dimen0 by 360\r@dian
		\r@duce
		\fi
	    \fi}

\def\Sine#1%
       {{%
	\dimen 0 = #1 \r@dian
	\r@duce
	\ifdim\dimen0 = -90\r@dian \then
	   \dimen4 = -1\r@dian
	   \c@mputefalse
	\fi
	\ifdim\dimen0 = 90\r@dian \then
	   \dimen4 = 1\r@dian
	   \c@mputefalse
	\fi
	\ifdim\dimen0 = 0\r@dian \then
	   \dimen4 = 0\r@dian
	   \c@mputefalse
	\fi
	\ifc@mpute \then
		\divide\dimen0 by 180
		\dimen0=3.141592654\dimen0
		\dimen 2 = 3.1415926535897963\r@dian 
		\divide\dimen 2 by 2 
		\Mess@ge {Sin: calculating Sin of \nodimen 0}%
		\count 0 = 1 
		\dimen 2 = 1 \r@dian 
		\dimen 4 = 0 \r@dian 
		\loop
			\ifnum	\dimen 2 = 0 
			\then	\stillc@nvergingfalse 
			\else	\stillc@nvergingtrue
			\fi
			\ifstillc@nverging 
			\then	\term {\count 0} {\dimen 0} {\dimen 2}%
				\advance \count 0 by 2
				\count 2 = \count 0
				\divide \count 2 by 2
				\ifodd	\count 2 
				\then	\advance \dimen 4 by \dimen 2
				\else	\advance \dimen 4 by -\dimen 2
				\fi
		\repeat
	\fi		
			\xdef \sine {\nodimen 4}%
       }}

\def\Cosine#1{\ifx\sine\UnDefined\edef\Savesine{\relax}\else
		             \edef\Savesine{\sine}\fi
	{\dimen0=#1\r@dian\advance\dimen0 by 90\r@dian
	 \Sine{\nodimen 0}
	 \xdef\cosine{\sine}
	 \xdef\sine{\Savesine}}}	      

\def\psdraft{
	\def\@psdraft{0}
}
\def\psfull{
	\def\@psdraft{100}
}

\psfull

\newif\if@scalefirst
\def\psscalefirst{\@scalefirsttrue}
\def\psrotatefirst{\@scalefirstfalse}
\psrotatefirst

\newif\if@draftbox
\def\psnodraftbox{
	\@draftboxfalse
}
\def\psdraftbox{
	\@draftboxtrue
}
\@draftboxtrue

\newif\if@prologfile
\newif\if@postlogfile
\def\pssilent{
	\@noisyfalse
}
\def\psnoisy{
	\@noisytrue
}
\psnoisy
\newif\if@bbllx
\newif\if@bblly
\newif\if@bburx
\newif\if@bbury
\newif\if@height
\newif\if@width
\newif\if@rheight
\newif\if@rwidth
\newif\if@angle
\newif\if@clip
\newif\if@verbose
\newif\if@scale
\def\@p@@sclip#1{\@cliptrue}

\newif\if@decmpr

\def\@p@@sfigure#1{\def\@p@sfile{null}\def\@p@sbbfile{null}
	        \openin1=#1.bb
		\ifeof1\closein1
	        	\openin1=\figurepath#1.bb
			\ifeof1\closein1
			        \openin1=#1
				\ifeof1\closein1%
				       \openin1=\figurepath#1
					\ifeof1
					   \ps@typeout{Error, File #1 not found}
						\if@bbllx\if@bblly
				   		\if@bburx\if@bbury
			      				\def\@p@sfile{#1}%
			      				\def\@p@sbbfile{#1}%
							\@decmprfalse
				  	   	\fi\fi\fi\fi
					\else\closein1
				    		\def\@p@sfile{\figurepath#1}%
				    		\def\@p@sbbfile{\figurepath#1}%
						\@decmprfalse
	                       		\fi%
			 	\else\closein1%
					\def\@p@sfile{#1}
					\def\@p@sbbfile{#1}
					\@decmprfalse
			 	\fi
			\else
				\def\@p@sfile{\figurepath#1}
				\def\@p@sbbfile{\figurepath#1.bb}
				\@decmprtrue
			\fi
		\else
			\def\@p@sfile{#1}
			\def\@p@sbbfile{#1.bb}
			\@decmprtrue
		\fi}

\def\@p@@sfile#1{\@p@@sfigure{#1}}

\def\@p@@sbbllx#1{
		\@bbllxtrue
		\dimen100=#1
		\edef\@p@sbbllx{\number\dimen100}
}
\def\@p@@sbblly#1{
		\@bbllytrue
		\dimen100=#1
		\edef\@p@sbblly{\number\dimen100}
}
\def\@p@@sbburx#1{
		\@bburxtrue
		\dimen100=#1
		\edef\@p@sbburx{\number\dimen100}
}
\def\@p@@sbbury#1{
		\@bburytrue
		\dimen100=#1
		\edef\@p@sbbury{\number\dimen100}
}
\def\@p@@sheight#1{
		\@heighttrue
		\dimen100=#1
   		\edef\@p@sheight{\number\dimen100}
}
\def\@p@@swidth#1{
		\@widthtrue
		\dimen100=#1
		\edef\@p@swidth{\number\dimen100}
}
\def\@p@@srheight#1{
		\@rheighttrue
		\dimen100=#1
		\edef\@p@srheight{\number\dimen100}
}
\def\@p@@srwidth#1{
		\@rwidthtrue
		\dimen100=#1
		\edef\@p@srwidth{\number\dimen100}
}
\def\@p@@sangle#1{
		\@angletrue
		\edef\@p@sangle{#1} 
}
\def\@p@@srotate#1{\@p@@sangle{-#1}}
\def\@p@@sscale#1{
		\@scaletrue
		\edef\@p@sscale{#1}
}
\def\@p@@ssilent#1{ 
		\@verbosefalse
}
\def\@p@@sprolog#1{\@prologfiletrue\def\@prologfileval{#1}}
\def\@p@@spostlog#1{\@postlogfiletrue\def\@postlogfileval{#1}}
\def\@cs@name#1{\csname #1\endcsname}
\def\@setparms#1=#2,{\@cs@name{@p@@s#1}{#2}}
\def\ps@init@parms{
		\@bbllxfalse \@bbllyfalse
		\@bburxfalse \@bburyfalse
		\@heightfalse \@widthfalse
		\@rheightfalse \@rwidthfalse
		\@scalefalse
		\def\@p@sbbllx{}\def\@p@sbblly{}
		\def\@p@sbburx{}\def\@p@sbbury{}
		\def\@p@sheight{}\def\@p@swidth{}
		\def\@p@srheight{}\def\@p@srwidth{}
		\def\@p@sangle{0}
		\def\@p@sfile{} \def\@p@sbbfile{}
		\def\@p@scost{10}
		\def\@sc{}
		\@prologfilefalse
		\@postlogfilefalse
		\@clipfalse
		\if@noisy
			\@verbosetrue
		\else
			\@verbosefalse
		\fi
}
\def\parse@ps@parms#1{
	 	\@psdo\@psfiga:=#1\do
		   {\expandafter\@setparms\@psfiga,}}
\newif\ifno@bb
\def\bb@missing{
	\if@verbose{
		\ps@typeout{psfig: searching \@p@sbbfile \space  for bounding box}
	}\fi
	\no@bbtrue
	\epsf@getbb{\@p@sbbfile}
        \ifno@bb \else \bb@cull\epsf@llx\epsf@lly\epsf@urx\epsf@ury\fi
}	
\def\bb@cull#1#2#3#4{
	\dimen100=#1 bp\edef\@p@sbbllx{\number\dimen100}
	\dimen100=#2 bp\edef\@p@sbblly{\number\dimen100}
	\dimen100=#3 bp\edef\@p@sbburx{\number\dimen100}
	\dimen100=#4 bp\edef\@p@sbbury{\number\dimen100}
	\no@bbfalse
}
\newdimen\p@intvaluex
\newdimen\p@intvaluey
\def\rotate@#1#2{{\dimen0=#1 sp\dimen1=#2 sp
		  \global\p@intvaluex=\cosine\dimen0
		  \dimen3=\sine\dimen1
		  \global\advance\p@intvaluex by -\dimen3
		  \global\p@intvaluey=\sine\dimen0
		  \dimen3=\cosine\dimen1
		  \global\advance\p@intvaluey by \dimen3
		  }}
\def\compute@bb{
		\no@bbfalse
		\if@bbllx \else \no@bbtrue \fi
		\if@bblly \else \no@bbtrue \fi
		\if@bburx \else \no@bbtrue \fi
		\if@bbury \else \no@bbtrue \fi
		\ifno@bb \bb@missing \fi
		\ifno@bb \ps@typeout{FATAL ERROR: no bb supplied or found}
			\no-bb-error
		\fi
		\count203=\@p@sbburx
		\count204=\@p@sbbury
		\advance\count203 by -\@p@sbbllx
		\advance\count204 by -\@p@sbblly
		\edef\ps@bbw{\number\count203}
		\edef\ps@bbh{\number\count204}
		\if@angle 
			\Sine{\@p@sangle}\Cosine{\@p@sangle}
	        	{\dimen100=\maxdimen\xdef\r@p@sbbllx{\number\dimen100}
					    \xdef\r@p@sbblly{\number\dimen100}
			                    \xdef\r@p@sbburx{-\number\dimen100}
					    \xdef\r@p@sbbury{-\number\dimen100}}
                        \def\minmaxtest{
			   \ifnum\number\p@intvaluex<\r@p@sbbllx
			      \xdef\r@p@sbbllx{\number\p@intvaluex}\fi
			   \ifnum\number\p@intvaluex>\r@p@sbburx
			      \xdef\r@p@sbburx{\number\p@intvaluex}\fi
			   \ifnum\number\p@intvaluey<\r@p@sbblly
			      \xdef\r@p@sbblly{\number\p@intvaluey}\fi
			   \ifnum\number\p@intvaluey>\r@p@sbbury
			      \xdef\r@p@sbbury{\number\p@intvaluey}\fi
			   }
			\rotate@{\@p@sbbllx}{\@p@sbblly}
			\minmaxtest
			\rotate@{\@p@sbbllx}{\@p@sbbury}
			\minmaxtest
			\rotate@{\@p@sbburx}{\@p@sbblly}
			\minmaxtest
			\rotate@{\@p@sbburx}{\@p@sbbury}
			\minmaxtest
			\edef\@p@sbbllx{\r@p@sbbllx}\edef\@p@sbblly{\r@p@sbblly}
			\edef\@p@sbburx{\r@p@sbburx}\edef\@p@sbbury{\r@p@sbbury}
		\fi
		\count203=\@p@sbburx
		\count204=\@p@sbbury
		\advance\count203 by -\@p@sbbllx
		\advance\count204 by -\@p@sbblly
		\edef\@bbw{\number\count203}
		\edef\@bbh{\number\count204}
}
\def\in@hundreds#1#2#3{\count240=#2 \count241=#3
		     \count100=\count240	
		     \divide\count100 by \count241
		     \count101=\count100
		     \multiply\count101 by \count241
		     \advance\count240 by -\count101
		     \multiply\count240 by 10
		     \count101=\count240	
		     \divide\count101 by \count241
		     \count102=\count101
		     \multiply\count102 by \count241
		     \advance\count240 by -\count102
		     \multiply\count240 by 10
		     \count102=\count240	
		     \divide\count102 by \count241
		     \count200=#1\count205=0
		     \count201=\count200
			\multiply\count201 by \count100
		 	\advance\count205 by \count201
		     \count201=\count200
			\divide\count201 by 10
			\multiply\count201 by \count101
			\advance\count205 by \count201
		     \count201=\count200
			\divide\count201 by 100
			\multiply\count201 by \count102
			\advance\count205 by \count201
		     \edef\@result{\number\count205}
}
\def\ps@scaleinhundreds#1{
		\in@hundreds{#1}{\@p@sscale}{100}
		\edef#1{\@result}
}
\def\compute@wfromh{
		\in@hundreds{\@p@sheight}{\@bbw}{\@bbh}
		\edef\@p@swidth{\@result}
}
\def\compute@hfromw{
	        \in@hundreds{\@p@swidth}{\@bbh}{\@bbw}
		\edef\@p@sheight{\@result}
}
\def\compute@handw{
		\if@height 
			\if@width
			\else
				\compute@wfromh
			\fi
		\else 
			\if@width
				\compute@hfromw
			\else
				\edef\@p@sheight{\@bbh}
				\edef\@p@swidth{\@bbw}
			\fi
		\fi
}
\def\compute@resv{
		\if@rheight \else \edef\@p@srheight{\@p@sheight} \fi
		\if@rwidth \else \edef\@p@srwidth{\@p@swidth} \fi
}
\def\compute@sizes{
	\compute@bb
	\if@scalefirst\if@angle
	\if@width
	   \in@hundreds{\@p@swidth}{\@bbw}{\ps@bbw}
	   \edef\@p@swidth{\@result}
	\fi
	\if@height
	   \in@hundreds{\@p@sheight}{\@bbh}{\ps@bbh}
	   \edef\@p@sheight{\@result}
	\fi
	\fi\fi
	\compute@handw
	\compute@resv
	\if@scale
	   \if@verbose
	      \ps@typeout{(scaling by \@p@sscale)}%
	   \fi
	   \ps@scaleinhundreds{\@p@swidth}%
	   \ps@scaleinhundreds{\@p@sheight}%
	   \ps@scaleinhundreds{\@p@srwidth}%
	   \ps@scaleinhundreds{\@p@srheight}%
	\fi
}

\def\psfig#1{\vbox {
	\ps@init@parms
	\parse@ps@parms{#1}
	\compute@sizes
	\ifnum\@p@scost<\@psdraft{
		\special{ps::[begin] 	\@p@swidth \space \@p@sheight \space
				\@p@sbbllx \space \@p@sbblly \space
				\@p@sbburx \space \@p@sbbury \space
				startTexFig \space }
		\if@angle
			\special {ps:: \@p@sangle \space rotate \space} 
		\fi
		\if@clip{
			\if@verbose{
				\ps@typeout{(clip)}
			}\fi
			\special{ps:: doclip \space }
		}\fi
		\if@prologfile
		    \special{ps: plotfile \@prologfileval \space } \fi
		\if@decmpr{
			\if@verbose{
				\ps@typeout{psfig: including \@p@sfile.Z \space }
			}\fi
			\special{ps: plotfile "`zcat \@p@sfile.Z" \space }
		}\else{
			\if@verbose{
				\ps@typeout{psfig: including \@p@sfile \space }
			}\fi
			\special{ps: plotfile \@p@sfile \space }
		}\fi
		\if@postlogfile
		    \special{ps: plotfile \@postlogfileval \space } \fi
		\special{ps::[end] endTexFig \space }
		\vbox to \@p@srheight true sp{
			\hbox to \@p@srwidth true sp{
				\hss
			}
		\vss
		}
	}\else{
		\if@draftbox{		
			\hbox{\frame{\vbox to \@p@srheight true sp{
			\vss
			\hbox to \@p@srwidth true sp{ \hss \@p@sfile \hss }
			\vss
			}}}
		}\else{
			\vbox to \@p@srheight true sp{
			\vss
			\hbox to \@p@srwidth true sp{\hss}
			\vss
			}
		}\fi

	}\fi
}}
\psfigRestoreAt
\let\@=\LaTeXAtSign